\documentclass[journal=jctcce,manuscript=article,layout=twocolumn]{achemso}

\usepackage[T1]{fontenc} 
\usepackage{hyperref}       
\usepackage{graphicx}       
\usepackage{url}            
\usepackage{booktabs}       
\usepackage{natbib}
\usepackage{standalone}
\usepackage{amsmath}
\usepackage{siunitx}
\usepackage{import}
\usepackage[draft]{minted}

\setminted{ xleftmargin=25pt, baselinestretch=1, bgcolor=grey, fontsize=\footnotesize}

\newcommand{\Vharmonicprior}{V^\mathrm{(prior)}_\mathrm{harmonic}}
\newcommand{\Vrepulsionprior}{V^\mathrm{(prior)}_\mathrm{repulsive}}

\author{Stefan Doerr}
\affiliation[Acellera]
{Acellera, Barcelona, Spain}
\author{Maciej Majewski}
\affiliation[UPF]
{Computational Science Laboratory, Universitat Pompeu Fabra, Barcelona, Spain}
\author{Adri\`a P\'erez}
\affiliation[UPF]
{Computational Science Laboratory, Universitat Pompeu Fabra, Barcelona, Spain}
\author{Andreas Krämer}
\affiliation[maths]{Department of Mathematics and Computer Science, Freie Universität, Berlin, Germany}
\author{Cecilia Clementi}
\affiliation[phys]{Department of Physics, Freie Universität, Berlin, Germany}
\alsoaffiliation[rise]{Department of Chemistry, Rice University, Houston, Texas}
\author{Frank Noe}
\affiliation[maths]{Department of Mathematics and Computer Science, Freie Universität, Berlin, Germany}
\alsoaffiliation[phys]{Department of Physics, Freie Universität, Berlin, Germany}
\alsoaffiliation[rise]{Department of Chemistry, Rice University, Houston, Texas}
\author{Toni Giorgino}
\affiliation[CNR]{Biophysics Institute, National Research Council (CNR-IBF), Italy}
\alsoaffiliation[UNIMI]{Department of Biosciences, Università degli Studi di Milano, Italy}
\author{Gianni De Fabritiis}
\email{g.defabritiis@gmail.com}
\affiliation[UPF]
{Computational Science Laboratory, Universitat Pompeu Fabra, Barcelona, Spain}
\alsoaffiliation[Acellera]
{Acellera, Barcelona, Spain}
\alsoaffiliation[ICREA]{Instituci\'o Catalana de Recerca i Estudis Avan\c{c}ats, Barcelona, Spain}

\title{TorchMD: A deep learning framework for molecular simulations}

\abbreviations{MD, ML, CG, RMSD}

\begin{document}


\begin{abstract}
Molecular dynamics simulations provide a mechanistic description of molecules by relying on empirical potentials. The quality and transferability of such potentials can be improved leveraging data-driven models derived with machine learning approaches. Here, we present  TorchMD, a framework for molecular simulations with mixed classical and machine learning potentials. All of force computations including bond, angle, dihedral, Lennard-Jones and Coulomb interactions are expressed as PyTorch arrays and operations. Moreover, TorchMD enables learning and simulating neural network potentials. We validate it using standard Amber all-atom simulations, learning an ab-initio potential, performing an end-to-end training and finally learning and simulating a coarse-grained model for protein folding. We believe that TorchMD provides a useful tool-set to support molecular simulations of machine learning potentials. Code and data are freely available at \url{github.com/torchmd}.
\end{abstract}

\maketitle

\section{Introduction}

Classical molecular dynamics (MD) is a compute-intensive technique that enables 
quantitative studies of molecular processes.  Of the 
possible modeling approaches, classical all-atom MD  represents
all of the atoms of a chosen system explicitly (including solvent) and accounts for inter-atomic
forces through classical bonded and non-bonded potentials. It  has seen remarkable
developments due to its fidelity and it has been applied with
success to problems such as conformational
changes, folding, binding, permeation, and many others. \cite{Schulten_discovery_2009} 
It has, however, two important
limitations: first, it needs extensive and carefully-optimized tables of inter-atomic
potentials (known as force fields)\cite{ponder_force_2003}; second, 
it is compute-intensive, and despite heroic efforts and progress in
accelerating MD codes \cite{martinez-rosell_drug_2017}, it
still struggles to reach the temporal scales 
of several important physiological processes. 

Machine learning (ML) potentials have become
especially attractive with the advent of deep neural network (DNN) 
architectures, which enable the example-driven definition of arbitrarily
complex functions and their derivatives. As such, DNNs  offer a very
promising avenue to embed fast-yet-accurate potential energy
functions in MD simulations, after training on large-scale 
databases  obtained from more expensive approaches. One particularly interesting feature of neural network potentials is that they can learn many-body interactions. 
The SchNet architecture \cite{schutt2017schnet, schutt2018schnet}, for instance, learns a set of features using continuous filter convolutions on a graph neural network and predicts the forces and energy of the system.  SchNet was originally used in quantum chemistry to predict energies of small molecules form their atomistic representations. A key feature of using SchNet is that the model is inherently transferable across molecular systems. 
More recently, this has been extended to learn a potential of mean force which involves averaging of a potential over some coarse-grained degrees of freedom \cite{ruza2020temperature, duvenaud2015convolutional,husic2020coarse, wang2019machine, nuske2019coarse, wang2020ensemble,ZhangHan2018_CG}, which however pose challenges in their parameterization \cite{wang2019coarse,D0CC03512B}. 
Indeed, molecular modeling on a more granular scale has been tackled by so-called coarse-graining (CG)
approaches before \cite{marrink_perspective_2013,machado_sirah_2019,Saunders2013,Izvekov2005,Noid2013,ClementiCOSB}, but it is particularly interesting in combination with DNNs.

Here, we introduce TorchMD, a molecular dynamics code built from
scratch to leverage the primitives of the 
ML library PyTorch \cite{NEURIPS2019_pytorch}. 
TorchMD enables the rapid prototyping and integration of machine-learned potentials
by extending the  bonded and non-bonded force terms commonly used in MD 
with DNN-based ones of arbitrary complexity. The two key points of TorchMD are that, being written in PyTorch, it is very easy to integrate other ML PyTorch models, like ab-initio neural network potentials (NNPs) \cite{gao_torchani_2020, schutt2018schnet} and machine learning coarse-grained potentials \cite{wang2019machine, husic2020coarse}. Secondly, TorchMD provides the capability to perform end-to-end differentiable simulations \cite{D0CC03512B,wang2020endtoend}, being differentiable on all of its parameters. Similarly, Jax\cite{jax2018github} was used  to perform end-to-end molecular simulations on Lennard-Jones systems\cite{schoenholz2019jax} and   for biomolecular systems as well\cite{timemachine}. 
Other efforts have tackled the  integration of MD codes with
DNN libraries, although in different  contexts. For all-atom models,
\citet{wang2020endtoend} demonstrated the use of graph networks to recover empirical 
atom types. Ab-initio QM-based training of potentials is being tackled by several groups, including
\citet{yao_tensormol-01_2018, schnetpack, gao_torchani_2020} but not using a differentiable PyTorch environment.

This paper provides an account of the  capabilities of TorchMD  
(Section \ref{sec:features}), highlighting the  functional forms supported
and an effective fitting strategy for data-driven DNN potentials. 
All of the TorchMD code, including
the CLN tutorial and the corresponding training data,  is open-source and
available at  \url{github.com/torchmd}.

\section{Methods} \label{sec:features} 

\subsection{TorchMD simulations}

TorchMD is, at first glance, a standard molecular dynamics code. It offers NVT ensemble simulations including a Langevin thermostat. Starting atomic velocities are derived from a Maxwell-Boltzmann distribution. Integration is done using the velocity Verlet algorithm. Long-range electrostatics are approximated using the reaction field method \cite{rfe}. TorchMD also supports simulations of periodic systems. Minimization is done using the L-BFGS algorithm. Because it is written in Python using PyTorch arrays, it is also very simple to modify and simulations can be run on any devices supported by PyTorch (CPU, GPU, TPU). However, unlike specialized MD codes \cite{harvey2009acemd} it is not  designed for speed. TorchMD uses chemical units consistent with classical MD codes such as ACEMD\cite{harvey2009acemd}, namely kcal/mol for energies, K for temperatures, g/mol for masses and \r{A} for distances.

\subsection{Analytical potentials}

TorchMD supports reading AMBER force field parameters through parmed \cite{parmed}. In addition to that, to allow for faster prototyping and development, it implements its own easy to read YAML-based force field format.
An example force field YAML file for the simulation of a water box is given in Figure \ref{fig:yaml}.
Currently, TorchMD missing features include hydrogen bond constraints and neighbour lists.

\begin{figure}[h!]
\centering
\begin{minted}{yaml}
atomtypes: [OT, HT]

bonds:
  (OT, HT):
    k0: 450.000
    req: 0.9572
  (HT, HT):
    k0: 0.000
    req: 1.5139

angles:
  (HT, OT, HT):
    k0: 55.000
    theta0: 104.52

lj:
  OT:
    sigma: 3.1507
    epsilon: -0.1521
  HT:
    sigma: 0.4000
    epsilon: -0.0460

electrostatics:
  OT:
    charge: -0.834
  HT:
    charge: 0.417

masses:
  OT: 15.9994
  HT: 1.008
\end{minted}
\caption{An example YAML forcef ield for water molecules}
\label{fig:yaml}
\end{figure}

TorchMD implements the functional form of the AMBER potential \cite{maier_ff14sb:_2015}. It offers all basic AMBER terms: harmonic bonds, angles, torsions and non-bonded Van der Waals and electrostatic energies.
The above potentials are implemented as follows. The bonded potential terms are calculated as
$$V_\mathit{bonded} = k_0 (r - r_{eq})^2$$
where $k_0$ is the force constant, $r$ the distance between the bonded atoms and $r_{eq}$ the equilibrium distance between them.

The angle terms are calculated as
$$V_\mathit{angle} = k_\theta (\theta - \theta_{eq})^2$$
where $\theta$ the angle between the three bonded atoms, $k_\theta$ is the
angular force constant, and $\theta_{eq}$ the equilibrium angle.

The torsion terms are calculated as
$$V_\mathit{torsion} = \sum_{n=1}^{n_{max}} k_n (1 + \cos(n \phi - \gamma))$$
where $\phi$ the dihedral angle between the four atoms, $\gamma$ is the phase
offset and $k_n$ the amplitude of the harmonic component of periodicity~$n$.

The non-bonded Van der Waals (VdW) terms are calculated as
$$V_\mathit{VdW} = \frac{A}{r^{12}}-\frac{B}{r^6}$$
where $A = 4 \epsilon \sigma^{12}$ and $B = 4 \epsilon \sigma^6$ with $\epsilon$ being the well depth of the interaction of two atoms and $\sigma$ the distance at which the energy is zero. The VdW potential also supports a cutoff by using a switching distance. Its energy is then obtained by multiplying the $V_\mathit{VdW}$ term with the  scaling factor
$$S = 1 - 6 x^5 + 15 x^4 -10 x^3$$
$$\mbox{with} \quad x = (r - r_s)/(r_c - r_s)$$
where $r_{s}$ is the switching distance and $r_{c}$ the cutoff distance.

Electrostatics without cutoff are implemented using the following potential.
$$V_\mathit{electrostatic} = k_{e} \frac{q_i q_j}{r}$$
where $k_{e} = \frac{1}{4 \pi \epsilon_0}$ is Coulomb's constant, $q_i$ and $q_j$ the charges of the two atoms and $r$ the distance between them. Electrostatics with cutoff are modified to use the reaction field method \cite{rfe} as follows
$$V_\mathit{electrostatic} = k_{e} q_i q_j\left(\frac{1}{r}+{k}_{\mathit{rf}}{r}^{2}-{c}_{\mathit{rf}}\right)$$
$${k}_{\mathit{rf}}=\left(\frac{1}{{r_\mathit{c}}^3}\right)\left(\frac{{\epsilon}_{\mathit{sol}}-1}{2{\epsilon}_{\mathit{sol}}+1}\right)$$
$${c}_{\mathit{rf}}=\left(\frac{1}{{r}_{\mathit{c}}}\right)\left(\frac{3{\epsilon}_{\mathit{sol}}}{2{\epsilon}_{\mathit{sol}}+1}\right)$$
where $r_c$ corresponds to the cutoff distance and $\epsilon_{sol}$ to the solvent dielectric constant.

In addition to the above, TorchMD also trivially allows the use of any other external potential $V_{ext}$ written in PyTorch  which takes atomic coordinates as input and outputs energy and forces.

Thus the total potential is calculated as
\begin{equation}
\begin{split}
V_\mathit{total} & =  \sum^{n_{bonds}} V_\mathit{bonded} + \sum^{n_{angles}} V_\mathit{angle} + 
\sum^{n_{torsions}} \!\!\! V_\mathit{torsion} \\
& + \sum_{i}^{n_{atoms}}\sum_{j<i}^{n_{atoms}} (V_\mathit{VdW} + V_\mathit{electrostatic}) \\
& + \sum^{n_{ext}} V_\mathit{ext}
\end{split}
\end{equation}

Since PyTorch offers automatic differentiation, there is no need to calculate analytical gradients from the forces. Forces can be obtained with a single autograd PyTorch call on the total energy of the system. Analytical gradients have been nevertheless implemented for all analytical AMBER potential terms for performance reasons. 

\subsection{Training machine learning potentials}

TorchMD provides a fully usable code for training neural network potentials in PyTorch called TorchMD-Net(\url{github.com/torchmd/torchmd-net}). Currently we are using a SchNet-based\cite{schutt2017schnet} model. However, it would be straightforward to derive the forces from non-parametric kernel methods like FCHL,\cite{christensen2020fchl} by providing a simple force calculator class, or other ML potentials. This object just takes as input the positions and box every timestep and returns the external energies and forces computed with the method of choice.

For the present work,  we take the featurization and atom-wise layer from SchNetPack\cite{schnetpack}, but rewrote entirely the training and inference parts. In particular, to allow training on multiple GPUs, the network is trained using the  PyTorch lightning framework \cite{falcon2019pytorch}. 
TorchMD can also run concurrently a set of identical simulations by just changing the random number generator seed,  arranging the neural network potential into a batch for speed, thus recovering, at least partially,  the efficiency of optimized molecular dynamics codes. For the QM9 dataset, we trained the model using a standard loss function using mean square error over the energies. For the coarse-grain model, training is performed using the bottom-up ``force matching'' approach, focused on reproducing thermodynamics of the system from atomistic simulations, as described in previous work\cite{wang2019machine, husic2020coarse}.

\section{Results} \label{sec:cln}

To demonstrate functionalities of TorchMD, here we present some application examples. Firstly, a set of typical MD use cases (water box, small peptide, protein and ligand) mainly to assess speed and energy conservation. Secondly, we validate the code on QM9, a dataset of 134k small molecule conformations with energies  \cite{ramakrishnan2014quantum}. In this case, however we cannot run any dynamical simulations as the dataset only presents ground state conformations of the molecules, so we are mainly validating the training. Then, we demonstrate end-to-end differentiable capabilities of TorchMD by recovering force field parameters from a short MD trajectory. Finally, we present a coarse-grained simulation of a miniprotein, chignolin,\cite{honda2008crystal} using NNP trained on all-atom MD simulation data.   
Here, we also describe how to produce a neural network-based coarse-grained model of chignolin, although the methods exposed are general to any other protein. A step-by-step example of training and simulating CG model is presented in the tutorial available in the \url{github.com/torchmd/torchmd-cg} repository.

\subsection{Simulations of all-atom systems and performance}

\begin{center}
\begin{table}
\begin{tabular}{ lccc } 
 \hline
 System & Atoms & TorchMD & ACEMD \\
 \hline
 Water & 291 & 6 min 56 s & 7 s \\ 
 Di-alanine & 688 & 8 min 44 s & 8 s \\ 
 Trypsin & 3,248 & 13 min 2 s & 16 s \\ 
 \hline
\end{tabular}
\caption{Performance comparison for 50,000 steps at 1 fs/timestep on different systems.}
\label{tab:performance}
\end{table}
\end{center}

The performance of TorchMD is compared against ACEMD3\cite{harvey2009acemd}, a high-performance molecular dynamics code. In Table \ref{tab:performance} we can see the three different test systems comprised of a simple periodic water box of 97 water molecules, alanine dipeptide and trypsin with the ligand benzamidine bound to it. As it can be seen, TorchMD is around 60 times slower on the test systems than ACEMD3 running on a TITAN V NVIDIA GPU. Most of the performance discrepancy can be attributed to the lack of neighbour lists for non-bonded interactions in TorchMD and is currently prohibitive for much larger systems as the pair distances cannot fit into GPU memory. This is not a strongly limiting factor for the CG simulations conducted in this paper as the number of beads remains relatively low for the test case. However, we believe that, with an upcoming implementation of neighbour lists, TorchMD can reach much better performance, albeit still slower than highly specialized codes as ACEMD3 due to the generic nature of PyTorch operations in addition with the PyTorch library overhead.

To evaluate the correctness of the TorchMD implementation of the AMBER force-field we compared it against OpenMM for 14 different systems ranging from ions, water boxes, small molecules to whole proteins, thus testing all the different force-field terms. In all 14 test cases the potential energy difference between OpenMM and TorchMD was lower than $10^{-3}$ kcal/mol when computed with the same parameters.
Energy conservation was validated with TorchMD by running a NVE simulation of a periodic water box for 1 ns with 1 fs timestep. Energy conservation normalized per degree of freedom  was calculated as $E_{total} / n_{dof}  R $ where $n_{dof}=870$ is the number of degrees of freedom of the system  and $R$ the ideal gas constant. We obtained a mean value of $1.1~ 10^{-5}$ K per degree of freedom showing a good energy conservation.

\subsection{Training validation on the QM9 dataset}

The network was trained using Adam optimizer\cite{ kingma2014adam} with a learning rate scheduler on multiple GPUs by using PyTorch Lightning \cite{falcon2019pytorch}. An example of the configuration file for QM9 training is presented in Figure \ref{fig:train_input}. We performed multiple trainings using TorchMD-Net with different amounts of training data (Fig.\ref{fig:training}). The learning rate scheduler was determined with a patience of $10$ on a validation subset of $5\%$ of all data chosen at random. The performance reported is for the randomly chosen test set. The linear shape of the test performance in the log-log scale demonstrate the correctness of the training \cite{vapnik2013nature}. With the current set of hyperparameters (Fig.\ref{fig:train_input}) we report a best performance of $10$ meV for 100,000 training points, marginally better than the reported best performance of SchNet for QM9 \cite{schnetpack}.  

\begin{figure}[htpb!]
\begin{minted}{yaml}
batch_size: 128
cutoff: 5.0
data: ./qm9.db
label: energy_U0
lr: 0.0004
lr_factor: 0.8
lr_patience: 10
early_stopping_patience: 30
model: schnet
max_z: 100
num_epochs: 500
num_filters: 256
num_gaussians: 25
num_interactions: 6
test_ratio: 0.1
val_ratio: 0.05
\end{minted}
\caption{An example of a training input file for training QM9}
\label{fig:train_input}
\end{figure}

\begin{figure}[htb!]
\centering
\includegraphics[width=8cm]{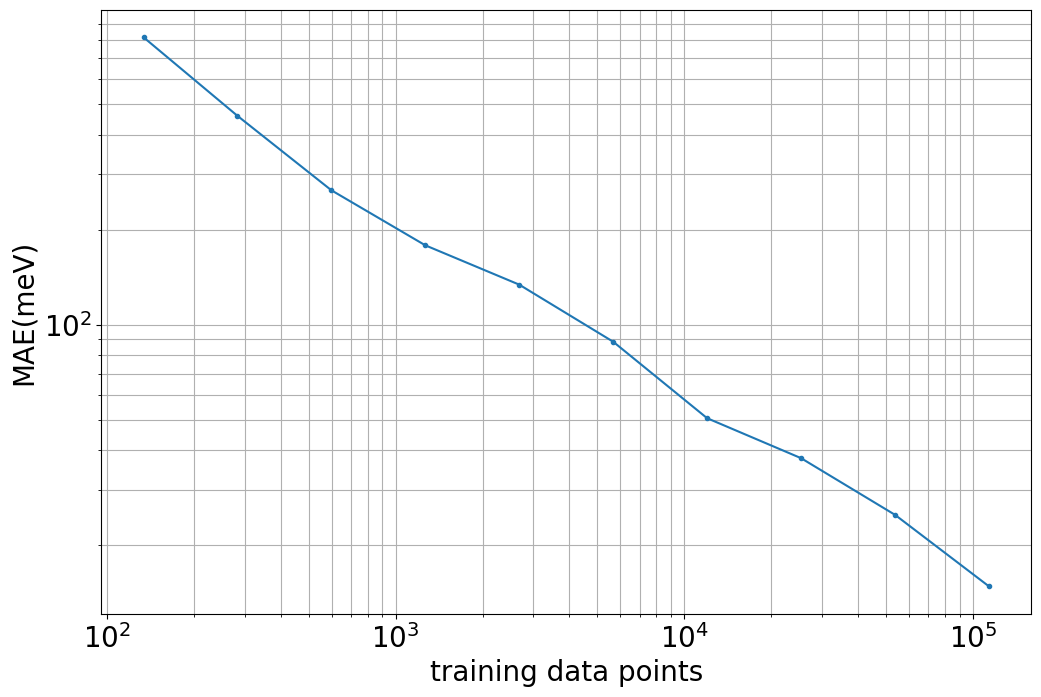}
\caption{Learning curve for the QM9 dataset.}
\label{fig:training}
\end{figure}

\subsection{Demonstration of end-to-end differentiable simulations}
The availability of automatic differentiation (AD) within a molecular dynamics package is beneficial beyond ML applications.
Being able to compute gradients for all numerical operations opens up new avenues for sensitivity analysis, force field optimization and steered MD simulations, as well as simulations under highly complex constraints and restraints. To demonstrate these capabilities, the present example infers force field parameters from a short MD trajectory.

First, a small water box containing 97 water molecules and one Na$^+$/Cl$^-$ ion pair was simulated using the TIP3P water model with flexible bonds and angles. After energy minimization and NVT equilibration at 300 K, the simulation was run for 10 ps in the microcanonical ensemble. The simulation used a 1 fs time step, a 9 \AA\ cutoff with 7.5 \AA\ switch distance, and reaction field electrostatics. Coordinates and velocities were saved every 10 steps.

Next, all partial atomic charges $q$ in the system were annihilated (in practice, they were scaled by 0.01 to ensure non-vanishing gradients of the electrostatic potential). In order to infer $q$ from the MD trajectory, the integrator was initialized with snapshots $r(t_i)$, $v(t_i)$ from the trajectory. Then, 10 steps of simulation were run with the modified charges and the final positions from this short simulation were compared with the respective subsequent trajectory snapshot $r(t_{i+1})$. In other words, the simulation served as a parameterized propagator
$Q: (r(t), v(t); q) \mapsto r(t + \delta_t)$ with $\delta_t = 10$ fs. Due to the AD capabilities within TorchMD, this propagator is end-to-end differentiable. 

\begin{figure}[htbp]
    \centering
    \includegraphics[width=\columnwidth]{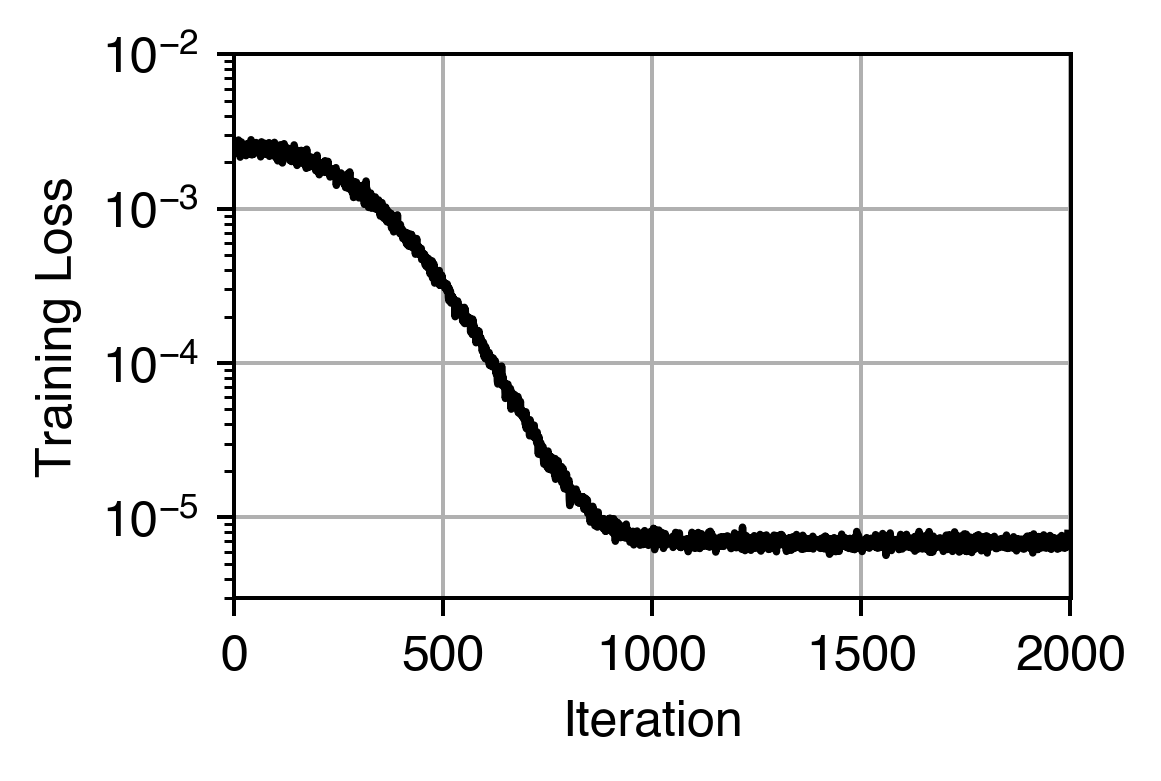}
    \includegraphics[width=\columnwidth]{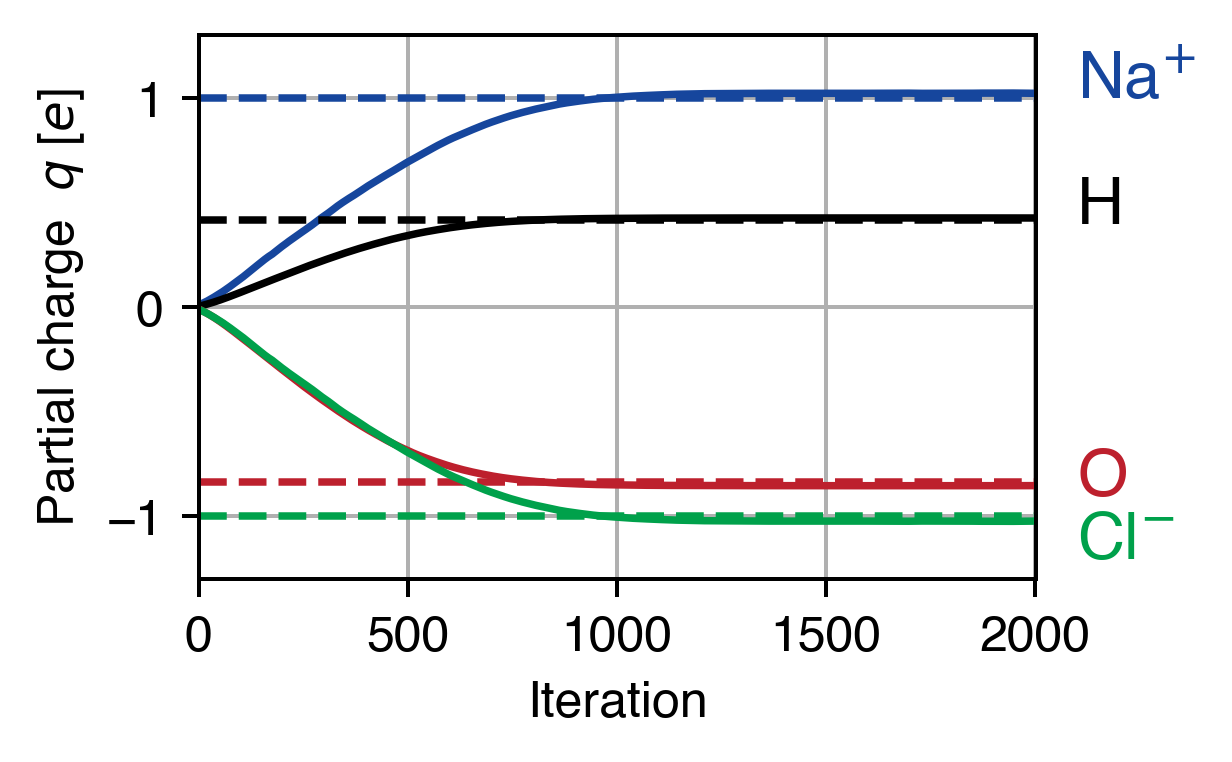}
    \caption{Inference of partial atomic charges $q$ from a short trajectory. Training loss (top) and charges (bottom) during training.}
    \label{fig:water}
\end{figure}

To recover the charges, we minimized the loss function
$$
L\big(r(t_i), v(t_i); q\big) = \| Q(r(t_i), v(t_i); q) - r(t_{i+1}) \|_2^2,
$$
i.e.\ the mean-squared distance between the ground-truth trajectory and the propagated coordinates (taking into account periodic boundary conditions). This loss function is differentiable with respect to the charges $q$ so that gradients can be obtained via backpropagation. Training was performed using Adam with a learning rate of $10^{-3}$ over one snapshot at a time. To enforce net neutrality, the positive charges ($q_H$ and $q_{\mathrm{Na}^+}$) were implicitly obtained from the oxygen and chlorine charges and only $q_O$ and $q_{\mathrm{Cl}^-}$ were explicitly optimized. 
Figure \ref{fig:water} shows the evolution of the training loss and the partial atomic charges during training. After just one epoch (1000 iterations), the original charges were recovered up to 3\% accuracy.

\subsection{Coarse-graining all-atom systems}

\begin{figure}[h!]
\centering
\includegraphics[width=\columnwidth]{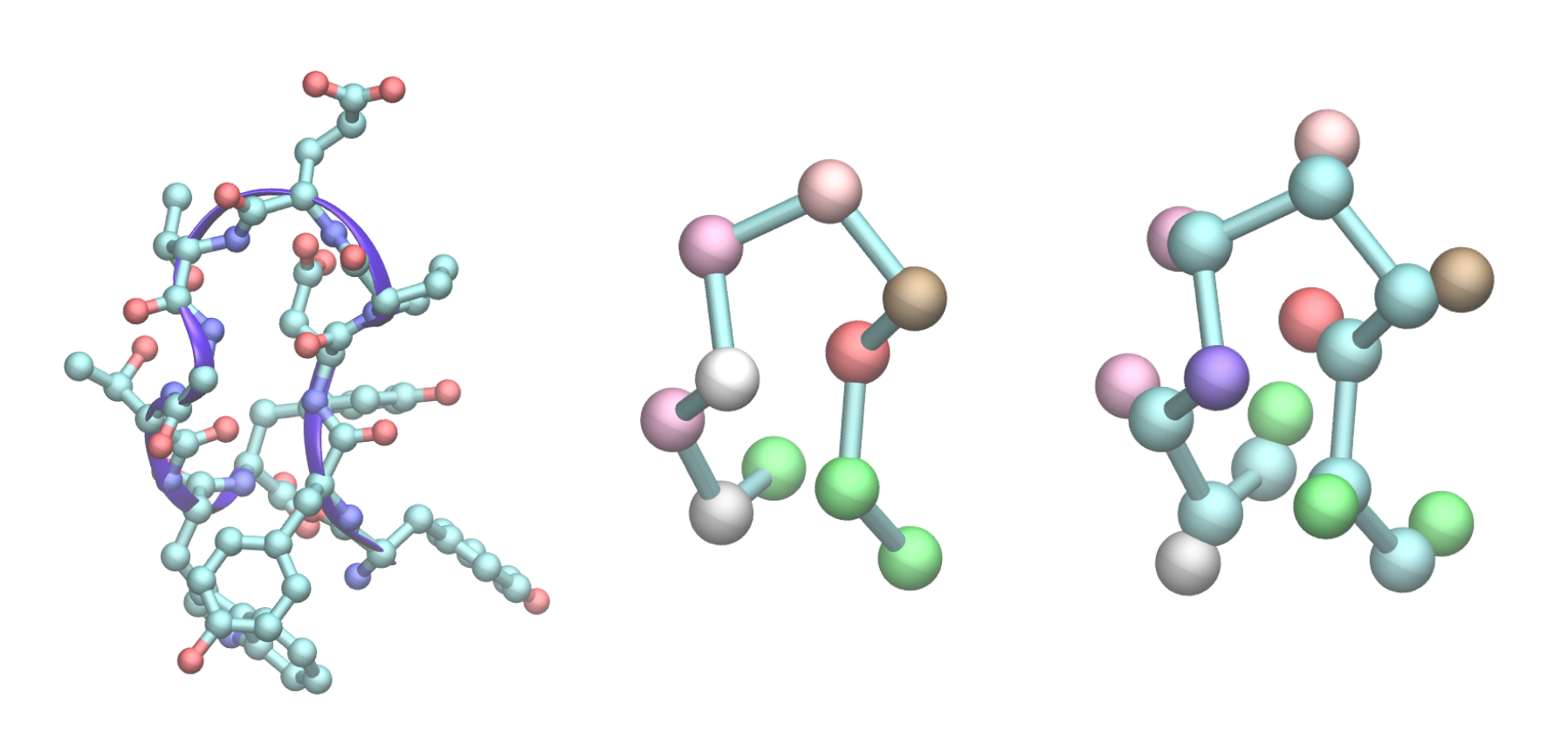}
\caption{Miniprotein chignolin: heavy-atom representation (left) and coarse-grained representations: CA beads connected by bonds (middle), CA and CB beads connected by bonds (right). The beads in coarse grained representations were coloured by bead type.}
\label{fig:cln-cg}
\end{figure}

For our last application example, we built two coarse-grained models of chignolin:  one solely based on $\alpha$-carbon atoms (CA) and  another one based on $\alpha$-carbon and $\beta$-carbon atoms (CACB) (Fig.\ref{fig:cln-cg}). 

\subsubsection{Training data}

We selected the CLN025 variant of chignolin (sequence YYDPETGTWY), which forms a $\beta$-hairpin turn while folded (Figure~\ref{fig:cln-cg}). Due to its small size (10 amino-acids) and fast folding, it has been extensively studied with MD \cite{lindorff2011fast, beauchamp2012simple, husic2016optimized, mckiernan2017modeling, sultan2018automated, scherer2019variational}. Training data was obtained from all-atom simulation of the protein in explicit solvent with ACEMD\cite{harvey2009acemd} on the GPUGRID.net distributed computing network.\cite{buch2010high} The system containing one chignolin chain was solvated in a cubic box of 40 \AA, containing 1881 water molecules and two Na$^+$ ions. The system was simulated at 350 K with CHARMM22* force field \cite{Piana2011} and TIP3P model of water.\cite{Jorgensen1983} A Langevin integrator was used with a damping constant of \SI{0.1}{\per\pico\second}. Integration time step was set to 4 fs, with heavy hydrogen atoms (scaled up to four times the hydrogen mass) and holonomic constrains on all hydrogen-heavy atom bond terms \cite{Feenstra1999}. Electrostatics were computed using Particle Mesh Ewald with a cutoff distance of 9 \AA\ and a grid spacing of 1 \AA. We used an adaptive sampling approach \cite{Doerr2014} where new simulations were started from the least explored states. As a result we obtained a total simulation time of \SI{180}{\us} with forces and coordinates saved every 100 ps giving a total of $1.8 \times 10^{6}$ frames. 

To obtain the training data for the CA model, the initial training set of coordinates and forces was filtered to retain only CA atoms positions and forces. In this example, a coarse grained system  contains 10 beads, built out of seven unique types of beads, one for each amino acid type. The training set for CACB model was prepared in a similar fashion, filtering both CA and CB atoms, and achieving 19 beads and 8 unique types of beads, as all CA atoms were classified as one bead type with the exception of glycine and each CB was assigned an amino acid-specific bead type. Details of bead selection for both models are described in Supporting Methods.

\subsubsection{Neural network potential training} \label{Pot_fit}

For coarse-grained simulations it is important to provide some prior potentials in order to limit the space that the dynamics can visit to the space sampled in the training data\cite{wang2019machine}. All the terms of the force-field could be applied, but  for simplicity we limit them to bonds and repulsions. Bonds prevent the protein polymer chain from breaking and repulsions stop computing NNP on very close atom distances where there is no data. 

For pairwise bonded terms, we used the all-atom training data to
construct distance histograms for each 
pair of bonded bead types. Specifically, for  each 
bonded pair, we counted the fraction of time 
that the respective distance spent in an equally-spaced bin in a distance range appropriate to the bead selection, 3.55~\AA\ and 4.2~\AA\ for all bonds between $\alpha$ carbon beads, and 1.3~\AA\ and~1.8 \AA\ for all bonds between $\alpha$ carbon and $\beta$ carbon. The distance distributions were Boltzmann-inverted 
to obtain free-energy profiles, and these were 
used to fit the equilibrium distance $r_0$ and the spring constant $k$ of the
respective harmonic potential $$ \Vharmonicprior(r) = k (r-r_0)^2 + V_0, $$
where $r$ is the distance between
the beads involved in the bond.

Priors for non-bonded repulsive terms were  derived analogously.
Distance histograms were constructed with 
30 equally-spaced distance bins between 3\,\AA\ and 6\,\AA\ and were used to 
fit the parameter $\epsilon$ of the repulsive potential
$$\Vrepulsionprior(r) = 4 \epsilon r^{-6} + V_0,$$
where $r$, as above, is the distance between the non-bonded beads.
In fitting the potential curves we corrected for the reference state  by normalizing counts of each bin by the volume of the corresponding spherical shell.
Non-linear curve fits were performed with the Levenberg-Marquardt 
method of the SciPy package \cite{virtanen_scipy_2020}.

The parameters of the prior forces are stored in a YAML force field file. Plots presenting the quality of fits are included in the Supporting Information (Fig. S1-S4) as well as YAML files describing prior force field. 

Based on resulting prior force field and input coordinates, we calculated a set of prior forces acting on the beads and then deducted them form true forces, resulting in a set of forces that we refer to as delta-forces. Along with coordinates, delta-forces were used as the input for training. 
In the case of CA model embeddings correspond to integers unique for each amino acid type. For CACB model all $\alpha$ carbons have the same embedding with the exception of glycine and each $\beta$ carbon has an embedding unique for each amino acid type. 

The network was trained using force matching approach, where a predicted force is compared to a true force from the training set \cite{wang2019machine, husic2020coarse}. 
In the example presented here, the network consisted of 3 interaction layers, 128 filters used in continuous-filter convolution, 128 features to describe atomic environments, 9 \AA\ cutoff radius and 150 Gaussian functions for CA model and $300$ Gaussians for CACB model. Increasing the number of Gaussian functions for CACB model was found to provide a higher stability of the model and prevent forming collapsed non-physical structures during the simulation. 
Models  for simulation were selected when the validation loss reached a plateau. The training and validation loss as well as learning rates are presented in Supporting Figure 5. 

\subsubsection{Simulation of the NNP}

\begin{figure}[h!]
\begin{minted}{yaml}
device: cuda:0
log_dir: sim_out_dir
output: output
topology: cln_ca_top.psf
coordinates: cln_ca_init_coords.xtc
replicas: 10
forcefield: cln_priors.yaml
forceterms:
- Bonds
- RepulsionCG
external:
  module: torchmdnet.nnp.calculators.torchmdcalc
  embeddings: [ 4,  4,  5,  8,  6, 13,  2, 13,  7,  4]
  file: train/epoch=80.ckpt
langevin_gamma: 1
langevin_temperature: 350
temperature: 350
precision: double
seed: 1
output_period: 1000
save_period: 1000
steps: 10000000
timestep: 1
\end{minted}
\caption{An example of a simulation input file}
\label{fig:input}
\end{figure}

\begin{figure*}[htpb!]
\centering
\includegraphics[width=\textwidth]{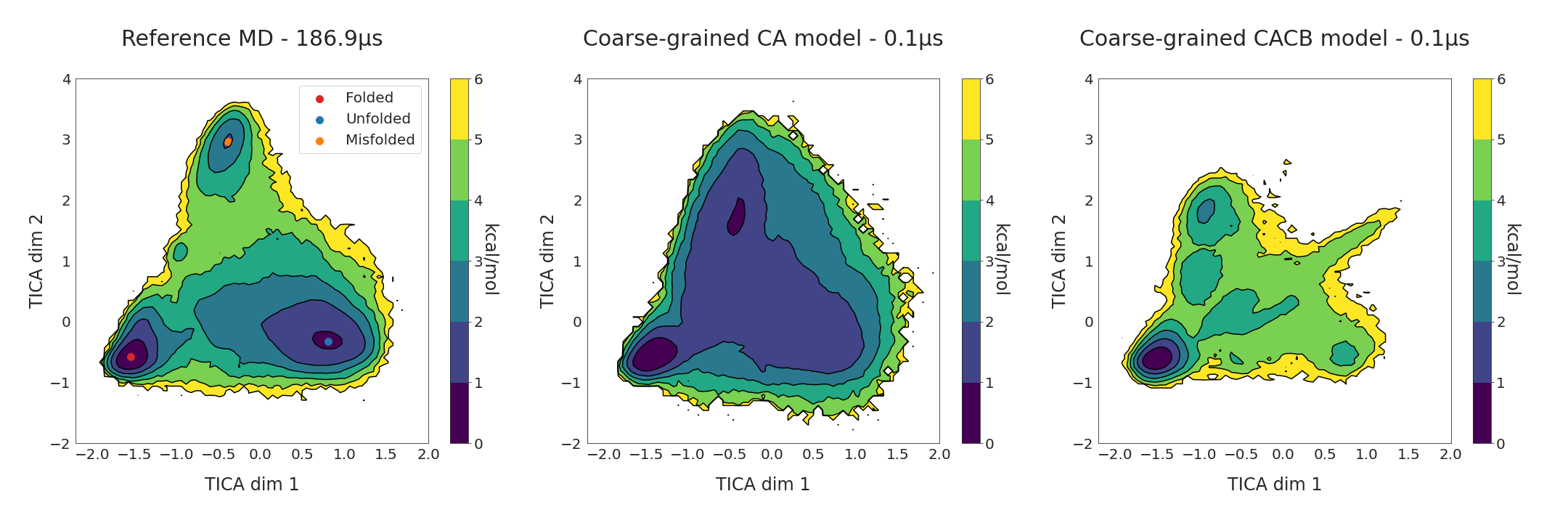}
\caption{Two-dimensional free energy surfaces for the reference all-atom MD simulations (left) and the two coarse-grained models, CA (center) and CACB (right). The free energy surface for each simulation set was obtained by binning over the first two TICA dimensions, dividing them into a 120 $\times$ 120 grid, and averaging the weights of the equilibrium probability in each bin computed by the Markov state model. The reference MD simulations plot displays the locations of the three energy minima on the surface, corresponding to folded state (red dot), unfolded conformations (blue dot) and a misfolded state (orange dot). Both reference MD and coarse-grained simulations were performed at 350K. 
}
\label{fig:Fig_TICA_CA}
\end{figure*}

The combinations of the force fields covering prior forces and the trained networks are used to simulate both CA and CACB systems with TorchMD. We introduce the parameters of the simulation as a YAML-formatted configuration file (Figure \ref{fig:input}), although the simulation can be also started from  command line. The network is introduced to TorchMD as an external force, with specified network's location, embeddings and a calculator. An  external force calculator class must have a "calculate" method that returns a tuple with energy and forces tensors. In our case, for both models, we run the simulation at 350K for 10 ns with 1 fs timestep, saving the output every 1000 fs. Note that while the simulations use a small timestep, the effective dynamics of the coarse-grained systems is much faster than the all-atom MD system as the coarse-grained model is supposed to reproduce the energetics but with much faster kinetics. Since TorchMD can easily handle parallel dynamics, we concurrently run ten simulations, of which five start from the folded state and five from unfolded conformations. 

\begin{figure*}[h!]
\centering
\includegraphics[width=\textwidth]{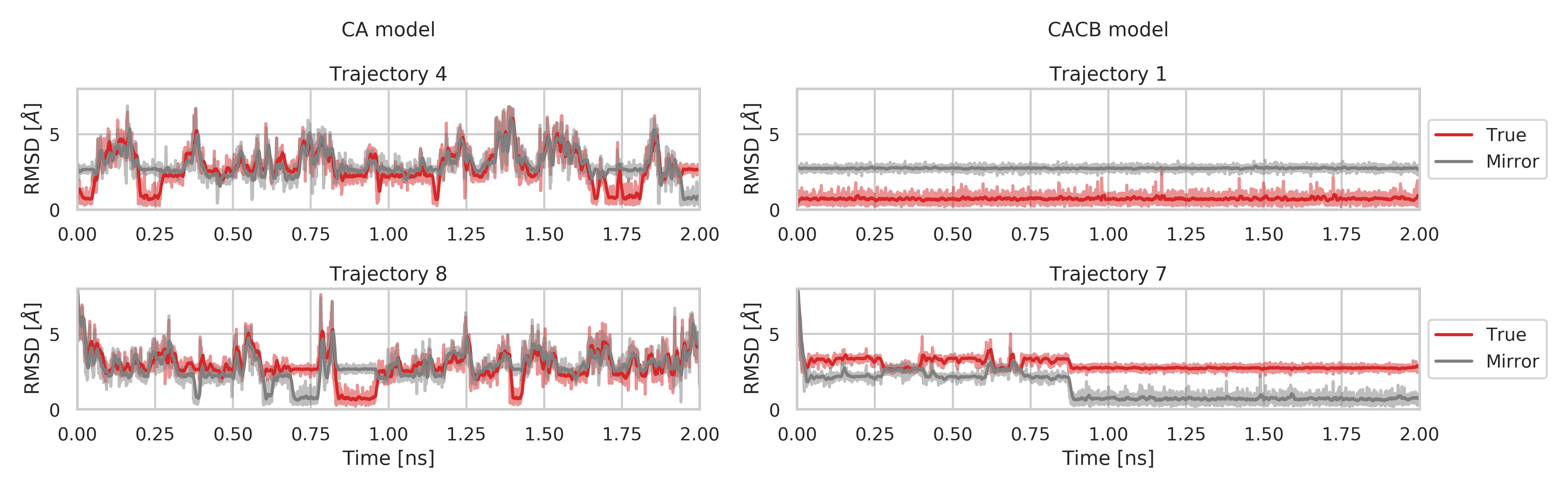}
\caption{The RMSD values across the first 2 ns of the unmodified trajectory (\emph{True}, red) and a mirror image of the original trajectory (\emph{Mirror}, gray) for CA model (on the left) and CACB model (on the right). Trajectory 4 (top left panel)  and Trajectory 1 (top right panel) are examples of a trajectories started from the folded state for CA model and CACB model, respectively. Trajectory 8 (bottom left panel) and Trajectory 7 (bottom right panel) are examples of trajectories started from elongated chain for CA model and CACB model, respectively. A moving average of 100 frames is represented as darker lines. The full 10 ns of each simulation is included in Supporting Figures S6-S9.}
\label{fig:Fig_RMSD_all}
\end{figure*}

The free energy surfaces obtained with a time-lagged independent component analysis (TICA)\cite{perez2013identification} for the all-atom baseline simulations and the coarse-grained simulations obtained with TorchMD are presented in Figure \ref{fig:Fig_TICA_CA}. The energy landscapes are obtained from binning the configurations over the first two TICA dimensions and computing the average of the equilibrium probability on each bin, obtained by Markov state model analysis of the microstate of each configuration. To support TICA plots, we included plots with RMSD values for the first 2 ns of representative trajectories for both models with different starting points (Fig.~\ref{fig:Fig_RMSD_all}). Plots presenting full trajectories are included in Supporting Information (Fig.S6-S9). Neither SchNet nor prior energy terms can enforce chirality in the system, because they both work purely on the distances between the beads. Therefore, the RMSD plots were supplemented with RMSD values of the trajectory's mirror image.

Results show that the coarse-grained simulations for both models were able to obtain several folding and unfolding events for chignolin. The energy landscapes for the CA model show that it captured the folded state as a global minimum of energy. The simulations also covers other minima representing unfolded and misfolded states. However, they do not recreate the energy barriers connecting these basins (as expected), which is better seen on the one-dimensional free energy surfaces (Fig.S10). The CACB model also detects the global minimum correctly, but fails at guessing the free energy of the unfolded region.  Overall the simulation is less stable than for CA model and the misfolded state minimum is incorrectly located.

\section{Conclusion}

In this paper we demonstrated TorchMD, a PyTorch-based molecular dynamics engine for biomolecular simulations with machine learning capabilities. We have shown several possible applications ranging from  Amber all-atom simulations, to end-to-end learning of parameters and finally a coarse-grained neural network potential for protein folding. In particular, building a NNP for protein folding requires supplementing it with asymptotic, analytical potentials for bonds and repulsions to prevent exploring conformations not visited in the training data in which the predictions of NNP are unreliable. We have shown how to coarse-grain a protein into either $\alpha$-carbon atoms or  $\alpha$-carbon and $\beta$-carbon atoms. Currently, CA model seems to work the best but future research will indicate which models are better suited for a more diverse set of targets.  TorchMD end-to-end differentiability of its parameters is a feature that projects such as the Open Force Field Initiative \cite{mobley_openforcefield_2018} can potentially exploit. Furthermore, for additional speed we plan to facilitate the integration of machine learning potentials in OpenMM\cite{eastman_openmm_2017} and ACEMD \cite{harvey2009acemd}. Meanwhile, we believe that TorchMD can play an important role by facilitating experimentation  between ML and MD fields,  speeding up the model-train-evaluate prototyping cycle, and promoting the adoption of data-based approaches in molecular simulations. All the code machinery to produce the models is made available for practitioners at \url{github.com/torchmd}.

\begin{acknowledgement}.
SD thanks the Chan Zuckerberg initiative for funding. We thank the volunteers
of  GPUGRID.net for donating computing time. This project has received funding from the European Union’s Horizon 2020 research and innovation programme under grant agreement No 823712.
\end{acknowledgement}


\bibliography{references.bib}

\providecommand{\latin}[1]{#1}
\makeatletter
\providecommand{\doi}
  {\begingroup\let\do\@makeother\dospecials
  \catcode`\{=1 \catcode`\}=2 \doi@aux}
\providecommand{\doi@aux}[1]{\endgroup\texttt{#1}}
\makeatother
\providecommand*\mcitethebibliography{\thebibliography}
\csname @ifundefined\endcsname{endmcitethebibliography}
  {\let\endmcitethebibliography\endthebibliography}{}
\begin{mcitethebibliography}{54}
\providecommand*\natexlab[1]{#1}
\providecommand*\mciteSetBstSublistMode[1]{}
\providecommand*\mciteSetBstMaxWidthForm[2]{}
\providecommand*\mciteBstWouldAddEndPuncttrue
  {\def\EndOfBibitem{\unskip.}}
\providecommand*\mciteBstWouldAddEndPunctfalse
  {\let\EndOfBibitem\relax}
\providecommand*\mciteSetBstMidEndSepPunct[3]{}
\providecommand*\mciteSetBstSublistLabelBeginEnd[3]{}
\providecommand*\EndOfBibitem{}
\mciteSetBstSublistMode{f}
\mciteSetBstMaxWidthForm{subitem}{(\alph{mcitesubitemcount})}
\mciteSetBstSublistLabelBeginEnd
  {\mcitemaxwidthsubitemform\space}
  {\relax}
  {\relax}

\bibitem[Lee \latin{et~al.}(2009)Lee, Hsin, Sotomayor, Comellas, and
  Schulten]{Schulten_discovery_2009}
Lee,~E.~H.; Hsin,~J.; Sotomayor,~M.; Comellas,~G.; Schulten,~K. Discovery
  {Through} the {Computational} {Microscope}. \emph{Structure} \textbf{2009},
  \emph{17}, 1295--1306\relax
\mciteBstWouldAddEndPuncttrue
\mciteSetBstMidEndSepPunct{\mcitedefaultmidpunct}
{\mcitedefaultendpunct}{\mcitedefaultseppunct}\relax
\EndOfBibitem
\bibitem[Ponder and Case(2003)Ponder, and Case]{ponder_force_2003}
Ponder,~J.~W.; Case,~D.~A. \emph{Advances in {Protein} {Chemistry}}; Protein
  {Simulations}; Academic Press, 2003; Vol.~66; pp 27--85\relax
\mciteBstWouldAddEndPuncttrue
\mciteSetBstMidEndSepPunct{\mcitedefaultmidpunct}
{\mcitedefaultendpunct}{\mcitedefaultseppunct}\relax
\EndOfBibitem
\bibitem[Martínez-Rosell \latin{et~al.}(2017)Martínez-Rosell, Giorgino,
  Harvey, and de~Fabritiis]{martinez-rosell_drug_2017}
Martínez-Rosell,~G.; Giorgino,~T.; Harvey,~M.~J.; de~Fabritiis,~G. Drug
  {Discovery} and {Molecular} {Dynamics}: {Methods}, {Applications} and
  {Perspective} {Beyond} the {Second} {Timescale}. \emph{Current Topics in
  Medicinal Chemistry} \textbf{2017}, \emph{17}, 2617--2625\relax
\mciteBstWouldAddEndPuncttrue
\mciteSetBstMidEndSepPunct{\mcitedefaultmidpunct}
{\mcitedefaultendpunct}{\mcitedefaultseppunct}\relax
\EndOfBibitem
\bibitem[Sch{\"u}tt \latin{et~al.}(2017)Sch{\"u}tt, Kindermans, Felix, Chmiela,
  Tkatchenko, and M{\"u}ller]{schutt2017schnet}
Sch{\"u}tt,~K.; Kindermans,~P.-J.; Felix,~H. E.~S.; Chmiela,~S.;
  Tkatchenko,~A.; M{\"u}ller,~K.-R. Schnet: A continuous-filter convolutional
  neural network for modeling quantum interactions. Advances in neural
  information processing systems. 2017; pp 991--1001\relax
\mciteBstWouldAddEndPuncttrue
\mciteSetBstMidEndSepPunct{\mcitedefaultmidpunct}
{\mcitedefaultendpunct}{\mcitedefaultseppunct}\relax
\EndOfBibitem
\bibitem[Sch{\"u}tt \latin{et~al.}(2018)Sch{\"u}tt, Sauceda, Kindermans,
  Tkatchenko, and M{\"u}ller]{schutt2018schnet}
Sch{\"u}tt,~K.~T.; Sauceda,~H.~E.; Kindermans,~P.-J.; Tkatchenko,~A.;
  M{\"u}ller,~K.-R. SchNet--A deep learning architecture for molecules and
  materials. \emph{The Journal of Chemical Physics} \textbf{2018}, \emph{148},
  241722\relax
\mciteBstWouldAddEndPuncttrue
\mciteSetBstMidEndSepPunct{\mcitedefaultmidpunct}
{\mcitedefaultendpunct}{\mcitedefaultseppunct}\relax
\EndOfBibitem
\bibitem[Ruza \latin{et~al.}(2020)Ruza, Wang, Schwalbe-Koda, Axelrod, Harris,
  and G{\'o}mez-Bombarelli]{ruza2020temperature}
Ruza,~J.; Wang,~W.; Schwalbe-Koda,~D.; Axelrod,~S.; Harris,~W.~H.;
  G{\'o}mez-Bombarelli,~R. Temperature-transferable coarse-graining of ionic
  liquids with dual graph convolutional neural networks. \emph{The Journal of
  Chemical Physics} \textbf{2020}, \emph{153}, 164501\relax
\mciteBstWouldAddEndPuncttrue
\mciteSetBstMidEndSepPunct{\mcitedefaultmidpunct}
{\mcitedefaultendpunct}{\mcitedefaultseppunct}\relax
\EndOfBibitem
\bibitem[Duvenaud \latin{et~al.}(2015)Duvenaud, Maclaurin, Iparraguirre,
  Bombarell, Hirzel, Aspuru-Guzik, and Adams]{duvenaud2015convolutional}
Duvenaud,~D.~K.; Maclaurin,~D.; Iparraguirre,~J.; Bombarell,~R.; Hirzel,~T.;
  Aspuru-Guzik,~A.; Adams,~R.~P. Convolutional networks on graphs for learning
  molecular fingerprints. Advances in neural information processing systems.
  2015; pp 2224--2232\relax
\mciteBstWouldAddEndPuncttrue
\mciteSetBstMidEndSepPunct{\mcitedefaultmidpunct}
{\mcitedefaultendpunct}{\mcitedefaultseppunct}\relax
\EndOfBibitem
\bibitem[Husic \latin{et~al.}(2020)Husic, Charron, Lemm, Wang, P{\'e}rez,
  Majewski, Kr{\"a}mer, Chen, Olsson, de~Fabritiis, No{\'e}, \latin{et~al.}
  others]{husic2020coarse}
Husic,~B.~E.; Charron,~N.~E.; Lemm,~D.; Wang,~J.; P{\'e}rez,~A.; Majewski,~M.;
  Kr{\"a}mer,~A.; Chen,~Y.; Olsson,~S.; de~Fabritiis,~G.; No{\'e},~F.,
  \latin{et~al.}  Coarse Graining Molecular Dynamics with Graph Neural
  Networks. \emph{J. Chem. Phys.} \textbf{2020}, \emph{153}, 194101\relax
\mciteBstWouldAddEndPuncttrue
\mciteSetBstMidEndSepPunct{\mcitedefaultmidpunct}
{\mcitedefaultendpunct}{\mcitedefaultseppunct}\relax
\EndOfBibitem
\bibitem[Wang \latin{et~al.}(2019)Wang, Olsson, Wehmeyer, P{\'e}rez, Charron,
  De~Fabritiis, No{\'e}, and Clementi]{wang2019machine}
Wang,~J.; Olsson,~S.; Wehmeyer,~C.; P{\'e}rez,~A.; Charron,~N.~E.;
  De~Fabritiis,~G.; No{\'e},~F.; Clementi,~C. Machine learning of
  coarse-grained molecular dynamics force fields. \emph{ACS central science}
  \textbf{2019}, \emph{5}, 755--767\relax
\mciteBstWouldAddEndPuncttrue
\mciteSetBstMidEndSepPunct{\mcitedefaultmidpunct}
{\mcitedefaultendpunct}{\mcitedefaultseppunct}\relax
\EndOfBibitem
\bibitem[N{\"u}ske \latin{et~al.}(2019)N{\"u}ske, Boninsegna, and
  Clementi]{nuske2019coarse}
N{\"u}ske,~F.; Boninsegna,~L.; Clementi,~C. Coarse-graining molecular systems
  by spectral matching. \emph{The Journal of chemical physics} \textbf{2019},
  \emph{151}, 044116\relax
\mciteBstWouldAddEndPuncttrue
\mciteSetBstMidEndSepPunct{\mcitedefaultmidpunct}
{\mcitedefaultendpunct}{\mcitedefaultseppunct}\relax
\EndOfBibitem
\bibitem[Wang \latin{et~al.}(2020)Wang, Chmiela, M{\"u}ller, No{\'e}, and
  Clementi]{wang2020ensemble}
Wang,~J.; Chmiela,~S.; M{\"u}ller,~K.-R.; No{\'e},~F.; Clementi,~C. Ensemble
  learning of coarse-grained molecular dynamics force fields with a kernel
  approach. \emph{The Journal of Chemical Physics} \textbf{2020}, \emph{152},
  194106\relax
\mciteBstWouldAddEndPuncttrue
\mciteSetBstMidEndSepPunct{\mcitedefaultmidpunct}
{\mcitedefaultendpunct}{\mcitedefaultseppunct}\relax
\EndOfBibitem
\bibitem[Zhang \latin{et~al.}(2018)Zhang, Han, Wang, Car, and
  E]{ZhangHan2018_CG}
Zhang,~L.; Han,~J.; Wang,~H.; Car,~R.; E,~W. {DeePCG}: constructing
  coarse-grained models via deep neural networks. \emph{arXiv:1802.08549}
  \textbf{2018}, \relax
\mciteBstWouldAddEndPunctfalse
\mciteSetBstMidEndSepPunct{\mcitedefaultmidpunct}
{}{\mcitedefaultseppunct}\relax
\EndOfBibitem
\bibitem[Wang and G{\'o}mez-Bombarelli(2019)Wang, and
  G{\'o}mez-Bombarelli]{wang2019coarse}
Wang,~W.; G{\'o}mez-Bombarelli,~R. Coarse-graining auto-encoders for molecular
  dynamics. \emph{npj Computational Materials} \textbf{2019}, \emph{5},
  1--9\relax
\mciteBstWouldAddEndPuncttrue
\mciteSetBstMidEndSepPunct{\mcitedefaultmidpunct}
{\mcitedefaultendpunct}{\mcitedefaultseppunct}\relax
\EndOfBibitem
\bibitem[Wang \latin{et~al.}(2020)Wang, Yang, Harris, and
  Gómez-Bombarelli]{D0CC03512B}
Wang,~W.; Yang,~T.; Harris,~W.~H.; Gómez-Bombarelli,~R. Active learning and
  neural network potentials accelerate molecular screening of ether-based
  solvate ionic liquids. \emph{Chem. Commun.} \textbf{2020}, \emph{56},
  8920--8923\relax
\mciteBstWouldAddEndPuncttrue
\mciteSetBstMidEndSepPunct{\mcitedefaultmidpunct}
{\mcitedefaultendpunct}{\mcitedefaultseppunct}\relax
\EndOfBibitem
\bibitem[Marrink and Tieleman(2013)Marrink, and
  Tieleman]{marrink_perspective_2013}
Marrink,~S.~J.; Tieleman,~D.~P. Perspective on the {Martini} model.
  \emph{Chemical Society Reviews} \textbf{2013}, \emph{42}, 6801--6822\relax
\mciteBstWouldAddEndPuncttrue
\mciteSetBstMidEndSepPunct{\mcitedefaultmidpunct}
{\mcitedefaultendpunct}{\mcitedefaultseppunct}\relax
\EndOfBibitem
\bibitem[Machado \latin{et~al.}(2019)Machado, Barrera, Klein, Sóñora, Silva,
  and Pantano]{machado_sirah_2019}
Machado,~M.~R.; Barrera,~E.~E.; Klein,~F.; Sóñora,~M.; Silva,~S.; Pantano,~S.
  The {SIRAH} 2.0 {Force} {Field}: {Altius}, {Fortius}, {Citius}. \emph{Journal
  of Chemical Theory and Computation} \textbf{2019}, \emph{15},
  2719--2733\relax
\mciteBstWouldAddEndPuncttrue
\mciteSetBstMidEndSepPunct{\mcitedefaultmidpunct}
{\mcitedefaultendpunct}{\mcitedefaultseppunct}\relax
\EndOfBibitem
\bibitem[Saunders and Voth(2013)Saunders, and Voth]{Saunders2013}
Saunders,~M.~G.; Voth,~G.~A. Coarse-Graining Methods for Computational Biology.
  \emph{Annu. Rev. Biophys.} \textbf{2013}, \emph{42}, 73--93\relax
\mciteBstWouldAddEndPuncttrue
\mciteSetBstMidEndSepPunct{\mcitedefaultmidpunct}
{\mcitedefaultendpunct}{\mcitedefaultseppunct}\relax
\EndOfBibitem
\bibitem[Izvekov and Voth(2005)Izvekov, and Voth]{Izvekov2005}
Izvekov,~S.; Voth,~G.~A. A Multiscale Coarse-Graining Method for Biomolecular
  Systems. \emph{J. Phys. Chem. B} \textbf{2005}, \emph{109}, 2469--2473\relax
\mciteBstWouldAddEndPuncttrue
\mciteSetBstMidEndSepPunct{\mcitedefaultmidpunct}
{\mcitedefaultendpunct}{\mcitedefaultseppunct}\relax
\EndOfBibitem
\bibitem[Noid(2013)]{Noid2013}
Noid,~W.~G. Perspective: Coarse-grained models for biomolecular systems.
  \emph{J. Chem. Phys.} \textbf{2013}, \emph{139}, 090901\relax
\mciteBstWouldAddEndPuncttrue
\mciteSetBstMidEndSepPunct{\mcitedefaultmidpunct}
{\mcitedefaultendpunct}{\mcitedefaultseppunct}\relax
\EndOfBibitem
\bibitem[Clementi(2008)]{ClementiCOSB}
Clementi,~C. Coarse-grained models of protein folding: Toy-models or predictive
  tools? \emph{Curr. Opin. Struct. Biol.} \textbf{2008}, \emph{18},
  10--15\relax
\mciteBstWouldAddEndPuncttrue
\mciteSetBstMidEndSepPunct{\mcitedefaultmidpunct}
{\mcitedefaultendpunct}{\mcitedefaultseppunct}\relax
\EndOfBibitem
\bibitem[Paszke \latin{et~al.}(2019)Paszke, Gross, Massa, Lerer, Bradbury,
  Chanan, Killeen, Lin, Gimelshein, Antiga, Desmaison, Kopf, Yang, DeVito,
  Raison, Tejani, Chilamkurthy, Steiner, Fang, Bai, and
  Chintala]{NEURIPS2019_pytorch}
Paszke,~A.; Gross,~S.; Massa,~F.; Lerer,~A.; Bradbury,~J.; Chanan,~G.;
  Killeen,~T.; Lin,~Z.; Gimelshein,~N.; Antiga,~L.; Desmaison,~A.; Kopf,~A.;
  Yang,~E.; DeVito,~Z.; Raison,~M.; Tejani,~A.; Chilamkurthy,~S.; Steiner,~B.;
  Fang,~L.; Bai,~J.; Chintala,~S. In \emph{Advances in Neural Information
  Processing Systems 32}; Wallach,~H., Larochelle,~H., Beygelzimer,~A.,
  d\textquotesingle Alch\'{e}-Buc,~F., Fox,~E., Garnett,~R., Eds.; Curran
  Associates, Inc., 2019; pp 8024--8035\relax
\mciteBstWouldAddEndPuncttrue
\mciteSetBstMidEndSepPunct{\mcitedefaultmidpunct}
{\mcitedefaultendpunct}{\mcitedefaultseppunct}\relax
\EndOfBibitem
\bibitem[Gao \latin{et~al.}(2020)Gao, Ramezanghorbani, Isayev, Smith, and
  Roitberg]{gao_torchani_2020}
Gao,~X.; Ramezanghorbani,~F.; Isayev,~O.; Smith,~J.; Roitberg,~A. {TorchANI}:
  {A} {Free} and {Open} {Source} {PyTorch} {Based} {Deep} {Learning}
  {Implementation} of the {ANI} {Neural} {Network} {Potentials}. \textbf{2020},
  Publisher: ChemRxiv\relax
\mciteBstWouldAddEndPuncttrue
\mciteSetBstMidEndSepPunct{\mcitedefaultmidpunct}
{\mcitedefaultendpunct}{\mcitedefaultseppunct}\relax
\EndOfBibitem
\bibitem[Wang \latin{et~al.}(2020)Wang, Fass, and Chodera]{wang2020endtoend}
Wang,~Y.; Fass,~J.; Chodera,~J.~D. End-to-End Differentiable Molecular
  Mechanics Force Field Construction. 2020; preprint arXiv:2010.01196\relax
\mciteBstWouldAddEndPuncttrue
\mciteSetBstMidEndSepPunct{\mcitedefaultmidpunct}
{\mcitedefaultendpunct}{\mcitedefaultseppunct}\relax
\EndOfBibitem
\bibitem[Bradbury \latin{et~al.}(2018)Bradbury, Frostig, Hawkins, Johnson,
  Leary, Maclaurin, and Wanderman-Milne]{jax2018github}
Bradbury,~J.; Frostig,~R.; Hawkins,~P.; Johnson,~M.~J.; Leary,~C.;
  Maclaurin,~D.; Wanderman-Milne,~S. {JAX}: composable transformations of
  {P}ython+{N}um{P}y programs. 2018; \url{http://github.com/google/jax}\relax
\mciteBstWouldAddEndPuncttrue
\mciteSetBstMidEndSepPunct{\mcitedefaultmidpunct}
{\mcitedefaultendpunct}{\mcitedefaultseppunct}\relax
\EndOfBibitem
\bibitem[Schoenholz and Cubuk(2019)Schoenholz, and Cubuk]{schoenholz2019jax}
Schoenholz,~S.~S.; Cubuk,~E.~D. JAX, M.D.: End-to-End Differentiable, Hardware
  Accelerated, Molecular Dynamics in Pure Python. 2019; preprint
  arXiv:1912.04232\relax
\mciteBstWouldAddEndPuncttrue
\mciteSetBstMidEndSepPunct{\mcitedefaultmidpunct}
{\mcitedefaultendpunct}{\mcitedefaultseppunct}\relax
\EndOfBibitem
\bibitem[Zhao(2020)]{timemachine}
Zhao,~Y. Time Machine. \emph{GitHub. Note:
  https://github.com/proteneer/timemachine} \textbf{2020}, \relax
\mciteBstWouldAddEndPunctfalse
\mciteSetBstMidEndSepPunct{\mcitedefaultmidpunct}
{}{\mcitedefaultseppunct}\relax
\EndOfBibitem
\bibitem[Yao \latin{et~al.}(2018)Yao, Herr, Toth, Mckintyre, and
  Parkhill]{yao_tensormol-01_2018}
Yao,~K.; Herr,~J.~E.; Toth,~D.~W.; Mckintyre,~R.; Parkhill,~J. The
  {TensorMol}-0.1 model chemistry: a neural network augmented with long-range
  physics. \emph{Chemical Science} \textbf{2018}, \emph{9}, 2261--2269,
  Publisher: The Royal Society of Chemistry\relax
\mciteBstWouldAddEndPuncttrue
\mciteSetBstMidEndSepPunct{\mcitedefaultmidpunct}
{\mcitedefaultendpunct}{\mcitedefaultseppunct}\relax
\EndOfBibitem
\bibitem[Schütt \latin{et~al.}(2019)Schütt, Kessel, Gastegger, Nicoli,
  Tkatchenko, and Müller]{schnetpack}
Schütt,~K.~T.; Kessel,~P.; Gastegger,~M.; Nicoli,~K.~A.; Tkatchenko,~A.;
  Müller,~K.-R. SchNetPack: A Deep Learning Toolbox For Atomistic Systems.
  \emph{Journal of Chemical Theory and Computation} \textbf{2019}, \emph{15},
  448--455\relax
\mciteBstWouldAddEndPuncttrue
\mciteSetBstMidEndSepPunct{\mcitedefaultmidpunct}
{\mcitedefaultendpunct}{\mcitedefaultseppunct}\relax
\EndOfBibitem
\bibitem[Tironi \latin{et~al.}(1995)Tironi, Sperb, Smith, and van
  Gunsteren]{rfe}
Tironi,~I.~G.; Sperb,~R.; Smith,~P.~E.; van Gunsteren,~W.~F. A generalized
  reaction field method for molecular dynamics simulations. \emph{The Journal
  of Chemical Physics} \textbf{1995}, \emph{102}, 5451--5459\relax
\mciteBstWouldAddEndPuncttrue
\mciteSetBstMidEndSepPunct{\mcitedefaultmidpunct}
{\mcitedefaultendpunct}{\mcitedefaultseppunct}\relax
\EndOfBibitem
\bibitem[Harvey \latin{et~al.}(2009)Harvey, Giupponi, and
  Fabritiis]{harvey2009acemd}
Harvey,~M.~J.; Giupponi,~G.; Fabritiis,~G.~D. ACEMD: accelerating biomolecular
  dynamics in the microsecond time scale. \emph{Journal of chemical theory and
  computation} \textbf{2009}, \emph{5}, 1632--1639\relax
\mciteBstWouldAddEndPuncttrue
\mciteSetBstMidEndSepPunct{\mcitedefaultmidpunct}
{\mcitedefaultendpunct}{\mcitedefaultseppunct}\relax
\EndOfBibitem
\bibitem[Shirts \latin{et~al.}(2017)Shirts, Klein, Swails, Yin, Gilson, Mobley,
  Case, and Zhong]{parmed}
Shirts,~M.~R.; Klein,~C.; Swails,~J.~M.; Yin,~J.; Gilson,~M.~K.; Mobley,~D.~L.;
  Case,~D.~A.; Zhong,~E.~D. Lessons learned from comparing molecular dynamics
  engines on the {SAMPL5} dataset. \emph{Journal of Computer-Aided Molecular
  Design} \textbf{2017}, \emph{31}, 147--161\relax
\mciteBstWouldAddEndPuncttrue
\mciteSetBstMidEndSepPunct{\mcitedefaultmidpunct}
{\mcitedefaultendpunct}{\mcitedefaultseppunct}\relax
\EndOfBibitem
\bibitem[Maier \latin{et~al.}(2015)Maier, Martinez, Kasavajhala, Wickstrom,
  Hauser, and Simmerling]{maier_ff14sb:_2015}
Maier,~J.~A.; Martinez,~C.; Kasavajhala,~K.; Wickstrom,~L.; Hauser,~K.~E.;
  Simmerling,~C. {ff14SB}: {Improving} the {Accuracy} of {Protein} {Side}
  {Chain} and {Backbone} {Parameters} from {ff99SB}. \emph{Journal of Chemical
  Theory and Computation} \textbf{2015}, \emph{11}, 3696--3713\relax
\mciteBstWouldAddEndPuncttrue
\mciteSetBstMidEndSepPunct{\mcitedefaultmidpunct}
{\mcitedefaultendpunct}{\mcitedefaultseppunct}\relax
\EndOfBibitem
\bibitem[Christensen \latin{et~al.}(2020)Christensen, Bratholm, Faber, and
  Anatole~von Lilienfeld]{christensen2020fchl}
Christensen,~A.~S.; Bratholm,~L.~A.; Faber,~F.~A.; Anatole~von Lilienfeld,~O.
  FCHL revisited: Faster and more accurate quantum machine learning. \emph{The
  Journal of Chemical Physics} \textbf{2020}, \emph{152}, 044107\relax
\mciteBstWouldAddEndPuncttrue
\mciteSetBstMidEndSepPunct{\mcitedefaultmidpunct}
{\mcitedefaultendpunct}{\mcitedefaultseppunct}\relax
\EndOfBibitem
\bibitem[Falcon(2019)]{falcon2019pytorch}
Falcon,~W. PyTorch Lightning. \emph{GitHub. Note:
  https://github.com/PyTorchLightning/pytorch-lightning} \textbf{2019}, \relax
\mciteBstWouldAddEndPunctfalse
\mciteSetBstMidEndSepPunct{\mcitedefaultmidpunct}
{}{\mcitedefaultseppunct}\relax
\EndOfBibitem
\bibitem[Ramakrishnan \latin{et~al.}(2014)Ramakrishnan, Dral, Rupp, and von
  Lilienfeld]{ramakrishnan2014quantum}
Ramakrishnan,~R.; Dral,~P.~O.; Rupp,~M.; von Lilienfeld,~O.~A. Quantum
  chemistry structures and properties of 134 kilo molecules. \emph{Scientific
  Data} \textbf{2014}, \emph{1}\relax
\mciteBstWouldAddEndPuncttrue
\mciteSetBstMidEndSepPunct{\mcitedefaultmidpunct}
{\mcitedefaultendpunct}{\mcitedefaultseppunct}\relax
\EndOfBibitem
\bibitem[Honda \latin{et~al.}(2008)Honda, Akiba, Kato, Sawada, Sekijima,
  Ishimura, Ooishi, Watanabe, Odahara, and Harata]{honda2008crystal}
Honda,~S.; Akiba,~T.; Kato,~Y.~S.; Sawada,~Y.; Sekijima,~M.; Ishimura,~M.;
  Ooishi,~A.; Watanabe,~H.; Odahara,~T.; Harata,~K. Crystal structure of a
  ten-amino acid protein. \emph{Journal of the American Chemical Society}
  \textbf{2008}, \emph{130}, 15327--15331\relax
\mciteBstWouldAddEndPuncttrue
\mciteSetBstMidEndSepPunct{\mcitedefaultmidpunct}
{\mcitedefaultendpunct}{\mcitedefaultseppunct}\relax
\EndOfBibitem
\bibitem[Kingma and Ba(2014)Kingma, and Ba]{kingma2014adam}
Kingma,~D.~P.; Ba,~J. Adam: A method for stochastic optimization. \emph{arXiv
  preprint arXiv:1412.6980} \textbf{2014}, \relax
\mciteBstWouldAddEndPunctfalse
\mciteSetBstMidEndSepPunct{\mcitedefaultmidpunct}
{}{\mcitedefaultseppunct}\relax
\EndOfBibitem
\bibitem[Vapnik(2013)]{vapnik2013nature}
Vapnik,~V. \emph{The nature of statistical learning theory}; Springer science
  \& business media, 2013\relax
\mciteBstWouldAddEndPuncttrue
\mciteSetBstMidEndSepPunct{\mcitedefaultmidpunct}
{\mcitedefaultendpunct}{\mcitedefaultseppunct}\relax
\EndOfBibitem
\bibitem[Lindorff-Larsen \latin{et~al.}(2011)Lindorff-Larsen, Piana, Dror, and
  Shaw]{lindorff2011fast}
Lindorff-Larsen,~K.; Piana,~S.; Dror,~R.~O.; Shaw,~D.~E. How fast-folding
  proteins fold. \emph{Science} \textbf{2011}, \emph{334}, 517--520\relax
\mciteBstWouldAddEndPuncttrue
\mciteSetBstMidEndSepPunct{\mcitedefaultmidpunct}
{\mcitedefaultendpunct}{\mcitedefaultseppunct}\relax
\EndOfBibitem
\bibitem[Beauchamp \latin{et~al.}(2012)Beauchamp, McGibbon, Lin, and
  Pande]{beauchamp2012simple}
Beauchamp,~K.~A.; McGibbon,~R.; Lin,~Y.-S.; Pande,~V.~S. Simple few-state
  models reveal hidden complexity in protein folding. \emph{Proceedings of the
  National Academy of Sciences} \textbf{2012}, \emph{109}, 17807--17813\relax
\mciteBstWouldAddEndPuncttrue
\mciteSetBstMidEndSepPunct{\mcitedefaultmidpunct}
{\mcitedefaultendpunct}{\mcitedefaultseppunct}\relax
\EndOfBibitem
\bibitem[Husic \latin{et~al.}(2016)Husic, McGibbon, Sultan, and
  Pande]{husic2016optimized}
Husic,~B.~E.; McGibbon,~R.~T.; Sultan,~M.~M.; Pande,~V.~S. Optimized parameter
  selection reveals trends in Markov state models for protein folding.
  \emph{The Journal of chemical physics} \textbf{2016}, \emph{145},
  194103\relax
\mciteBstWouldAddEndPuncttrue
\mciteSetBstMidEndSepPunct{\mcitedefaultmidpunct}
{\mcitedefaultendpunct}{\mcitedefaultseppunct}\relax
\EndOfBibitem
\bibitem[McKiernan \latin{et~al.}(2017)McKiernan, Husic, and
  Pande]{mckiernan2017modeling}
McKiernan,~K.~A.; Husic,~B.~E.; Pande,~V.~S. Modeling the mechanism of CLN025
  beta-hairpin formation. \emph{The Journal of chemical physics} \textbf{2017},
  \emph{147}, 104107\relax
\mciteBstWouldAddEndPuncttrue
\mciteSetBstMidEndSepPunct{\mcitedefaultmidpunct}
{\mcitedefaultendpunct}{\mcitedefaultseppunct}\relax
\EndOfBibitem
\bibitem[Sultan and Pande(2018)Sultan, and Pande]{sultan2018automated}
Sultan,~M.~M.; Pande,~V.~S. Automated design of collective variables using
  supervised machine learning. \emph{The Journal of chemical physics}
  \textbf{2018}, \emph{149}, 094106\relax
\mciteBstWouldAddEndPuncttrue
\mciteSetBstMidEndSepPunct{\mcitedefaultmidpunct}
{\mcitedefaultendpunct}{\mcitedefaultseppunct}\relax
\EndOfBibitem
\bibitem[Scherer \latin{et~al.}(2019)Scherer, Husic, Hoffmann, Paul, Wu, and
  No{\'e}]{scherer2019variational}
Scherer,~M.~K.; Husic,~B.~E.; Hoffmann,~M.; Paul,~F.; Wu,~H.; No{\'e},~F.
  Variational selection of features for molecular kinetics. \emph{The Journal
  of chemical physics} \textbf{2019}, \emph{150}, 194108\relax
\mciteBstWouldAddEndPuncttrue
\mciteSetBstMidEndSepPunct{\mcitedefaultmidpunct}
{\mcitedefaultendpunct}{\mcitedefaultseppunct}\relax
\EndOfBibitem
\bibitem[Buch \latin{et~al.}(2010)Buch, Harvey, Giorgino, Anderson, and
  De~Fabritiis]{buch2010high}
Buch,~I.; Harvey,~M.~J.; Giorgino,~T.; Anderson,~D.~P.; De~Fabritiis,~G.
  High-throughput all-atom molecular dynamics simulations using distributed
  computing. \emph{Journal of chemical information and modeling} \textbf{2010},
  \emph{50}, 397--403\relax
\mciteBstWouldAddEndPuncttrue
\mciteSetBstMidEndSepPunct{\mcitedefaultmidpunct}
{\mcitedefaultendpunct}{\mcitedefaultseppunct}\relax
\EndOfBibitem
\bibitem[Piana \latin{et~al.}(2011)Piana, Lindorff-Larsen, and Shaw]{Piana2011}
Piana,~S.; Lindorff-Larsen,~K.; Shaw,~D.~E. {How robust are protein folding
  simulations with respect to force field parameterization?} \emph{Biophysical
  Journal} \textbf{2011}, \emph{100}\relax
\mciteBstWouldAddEndPuncttrue
\mciteSetBstMidEndSepPunct{\mcitedefaultmidpunct}
{\mcitedefaultendpunct}{\mcitedefaultseppunct}\relax
\EndOfBibitem
\bibitem[Jorgensen \latin{et~al.}(1983)Jorgensen, Chandrasekhar, Madura, Impey,
  and Klein]{Jorgensen1983}
Jorgensen,~W.~L.; Chandrasekhar,~J.; Madura,~J.~D.; Impey,~R.~W.; Klein,~M.~L.
  {Comparison of simple potential functions for simulating liquid water}.
  \emph{J. Chem. Phys.} \textbf{1983}, \emph{79}, 926--935\relax
\mciteBstWouldAddEndPuncttrue
\mciteSetBstMidEndSepPunct{\mcitedefaultmidpunct}
{\mcitedefaultendpunct}{\mcitedefaultseppunct}\relax
\EndOfBibitem
\bibitem[Feenstra \latin{et~al.}(1999)Feenstra, Hess, and
  Berendsen]{Feenstra1999}
Feenstra,~K.~A.; Hess,~B.; Berendsen,~H.~J. {Improving efficiency of large
  time-scale molecular dynamics simulations of hydrogen-rich systems}. \emph{J.
  Comput. Chem.} \textbf{1999}, \emph{20}, 786--798\relax
\mciteBstWouldAddEndPuncttrue
\mciteSetBstMidEndSepPunct{\mcitedefaultmidpunct}
{\mcitedefaultendpunct}{\mcitedefaultseppunct}\relax
\EndOfBibitem
\bibitem[Doerr and {De Fabritiis}(2014)Doerr, and {De Fabritiis}]{Doerr2014}
Doerr,~S.; {De Fabritiis},~G. {On-the-Fly Learning and Sampling of Ligand
  Binding by High- Throughput Molecular Simulations}. \emph{Journal of Chemical
  Theory and Computation} \textbf{2014}, \relax
\mciteBstWouldAddEndPunctfalse
\mciteSetBstMidEndSepPunct{\mcitedefaultmidpunct}
{}{\mcitedefaultseppunct}\relax
\EndOfBibitem
\bibitem[Virtanen \latin{et~al.}(2020)Virtanen, Gommers, Oliphant, Haberland,
  Reddy, Cournapeau, Burovski, Peterson, Weckesser, Bright, van~der Walt,
  Brett, Wilson, Millman, Mayorov, Nelson, Jones, Kern, Larson, Carey, Polat,
  Feng, Moore, VanderPlas, Laxalde, Perktold, Cimrman, Henriksen, Quintero,
  Harris, Archibald, Ribeiro, Pedregosa, and van Mulbregt]{virtanen_scipy_2020}
Virtanen,~P.; Gommers,~R.; Oliphant,~T.~E.; Haberland,~M.; Reddy,~T.;
  Cournapeau,~D.; Burovski,~E.; Peterson,~P.; Weckesser,~W.; Bright,~J.;
  van~der Walt,~S.~J.; Brett,~M.; Wilson,~J.; Millman,~K.~J.; Mayorov,~N.;
  Nelson,~A. R.~J.; Jones,~E.; Kern,~R.; Larson,~E.; Carey,~C.~J.; Polat,~I.;
  Feng,~Y.; Moore,~E.~W.; VanderPlas,~J.; Laxalde,~D.; Perktold,~J.;
  Cimrman,~R.; Henriksen,~I.; Quintero,~E.~A.; Harris,~C.~R.; Archibald,~A.~M.;
  Ribeiro,~A.~H.; Pedregosa,~F.; van Mulbregt,~P. {SciPy} 1.0: fundamental
  algorithms for scientific computing in {Python}. \emph{Nature Methods}
  \textbf{2020}, \emph{17}, 261--272\relax
\mciteBstWouldAddEndPuncttrue
\mciteSetBstMidEndSepPunct{\mcitedefaultmidpunct}
{\mcitedefaultendpunct}{\mcitedefaultseppunct}\relax
\EndOfBibitem
\bibitem[P{\'e}rez-Hern{\'a}ndez \latin{et~al.}(2013)P{\'e}rez-Hern{\'a}ndez,
  Paul, Giorgino, De~Fabritiis, and No{\'e}]{perez2013identification}
P{\'e}rez-Hern{\'a}ndez,~G.; Paul,~F.; Giorgino,~T.; De~Fabritiis,~G.;
  No{\'e},~F. Identification of slow molecular order parameters for Markov
  model construction. \emph{The Journal of chemical physics} \textbf{2013},
  \emph{139}, 07B604\_1\relax
\mciteBstWouldAddEndPuncttrue
\mciteSetBstMidEndSepPunct{\mcitedefaultmidpunct}
{\mcitedefaultendpunct}{\mcitedefaultseppunct}\relax
\EndOfBibitem
\bibitem[Mobley \latin{et~al.}(2018)Mobley, Bannan, Rizzi, Bayly, Chodera, Lim,
  Lim, Beauchamp, Shirts, Gilson, and Eastman]{mobley_openforcefield_2018}
Mobley,~D.~L.; Bannan,~C.~C.; Rizzi,~A.; Bayly,~C.~I.; Chodera,~J.~D.;
  Lim,~V.~T.; Lim,~N.~M.; Beauchamp,~K.~A.; Shirts,~M.~R.; Gilson,~M.~K.;
  Eastman,~P.~K. Open {Force} {Field} {Consortium}: {Escaping} atom types using
  direct chemical perception with {SMIRNOFF} v0.1. 2018; Biorxiv preprint
  10.1101/286542\relax
\mciteBstWouldAddEndPuncttrue
\mciteSetBstMidEndSepPunct{\mcitedefaultmidpunct}
{\mcitedefaultendpunct}{\mcitedefaultseppunct}\relax
\EndOfBibitem
\bibitem[Eastman \latin{et~al.}(2017)Eastman, Swails, Chodera, McGibbon, Zhao,
  Beauchamp, Wang, Simmonett, Harrigan, Stern, Wiewiora, Brooks, and
  Pande]{eastman_openmm_2017}
Eastman,~P.; Swails,~J.; Chodera,~J.~D.; McGibbon,~R.~T.; Zhao,~Y.;
  Beauchamp,~K.~A.; Wang,~L.-P.; Simmonett,~A.~C.; Harrigan,~M.~P.;
  Stern,~C.~D.; Wiewiora,~R.~P.; Brooks,~B.~R.; Pande,~V.~S. {OpenMM} 7:
  {Rapid} development of high performance algorithms for molecular dynamics.
  \emph{PLOS Computational Biology} \textbf{2017}, \emph{13}, e1005659\relax
\mciteBstWouldAddEndPuncttrue
\mciteSetBstMidEndSepPunct{\mcitedefaultmidpunct}
{\mcitedefaultendpunct}{\mcitedefaultseppunct}\relax
\EndOfBibitem
\end{mcitethebibliography}

\end{document}